\documentstyle[mnras_cite,psfig]{mn}
\def\gsim{~\rlap{$>$}{\lower 1.0ex\hbox{$\sim$}}}
\def\lsim{~\rlap{$<$}{\lower 1.0ex\hbox{$\sim$}}}

\def\d{{\rm d}}
\def\PS{PhS}
\def\TL{TiL}

\begin{document}

\title[Environmental Dependence in the Luminosity Function]{The Effects of Photoionization on Galaxy Formation --- III:
Environmental Dependence in the Luminosity Function}
\author[A.~J.~Benson, C.~S.~Frenk, C.~M.~Baugh, S.~Cole \& C.~G.~Lacey]{A.~J.~Benson$^1$, C.~S.~Frenk$^2$, C.~M.~Baugh$^2$, S.~Cole$^2$ \& C.~G.~Lacey$^2$\\
1. California Institute of Technology, MC 105-24, Pasadena, CA 91125,
U.S.A. (e-mail: abenson@astro.caltech.edu) \\
2. Physics Department, University of Durham, Durham, DH1 3LE, England}

\maketitle

\begin{abstract}
Using semi-analytic modeling techniques, we calculate the luminosity
function of galaxy populations residing in cold dark matter halos of
different mass. We pay particular attention to the influence of the
reionization of the Universe on the number of faint galaxies and to
the effects of dynamical friction and tidal limitation of satellites
on the number of bright galaxies. We find substantial differences in
the shapes of the galaxy luminosity functions in halos of different
mass which reflect generic features of the cold dark matter model of
galaxy formation and thus offer the opportunity to test it. We then
consider how the individual halo luminosity functions combine together
to produce the global luminosity function.  Surprisingly, the global
function ends up having a shallower faint end slope than those of the
constituent halo luminosity functions. We compare our model
predictions with the limited datasets compiled by \scite{trenhodg}. We
find good agreement with the luminosity functions measured in the
Virgo and Coma clusters but significant disagreement with the
luminosity functions measured in the Local Group and Ursa Minor
cluster. We speculate on possible inadequacies in our modeling and in
the existing observational samples. The luminosity functions of
galaxies in groups and clusters identified in the 2dF and SDSS galaxy
redshift surveys offer the prospect of testing galaxy formation models
in detail.
\end{abstract}

\section{Introduction}

Understanding the galaxy luminosity function has been a goal of galaxy
formation theory for several decades
(e.g. \pcite{wr78,cole91,wf91}). A particularly interesting question
is whether the luminosity function is universal or whether it depends
on environmental factors such as the mass of the dark halo that hosts
a particular galaxy population.  Considerable attention has been paid
to the faint end of the luminosity function which has a much flatter
slope than the low mass end of the halo mass function predicted in
cold dark matter (CDM) models of galaxy formation (e.g. \pcite{norberg02}).

The early work of \scite{wr78} showed that the number of faint
galaxies must have been strongly affected by feedback processes that
prevented most of the gas from cooling in small halos at early times.
Some likely feedback mechanisms, such as the injection of energy into
the interstellar medium in the course of stellar evolution, depend on
the internal properties of the galaxy and so their effects may be
expected to be independent of the large-scale environment. A number of
observational studies, such as a recent analysis of the 2dF galaxy
redshift survey \cite{deprop02}, indeed find no significant difference
between the luminosity functions of galaxies in rich clusters and in
the field. Other studies, however, have found the opposite. For
example, \scite{phillipps87} concluded that galaxies in rich clusters
have luminosity functions with considerably steeper faint ends than
galaxies in the field. More recently, \scite{trenhodg} have claimed
that the faint end of the galaxy luminosity function varies
systematically with environment, increasing in slope from small,
diffuse systems like the Local Group, to massive, dense systems like
the Coma cluster.

In the cold dark matter model of galaxy formation, dark matter halos
retain considerable substructure after they collapse and virialize
(e.g. \pcite{klypin99,moore99}) and some of these subhalos are
associated with sites of galaxy formation. The mass function of
subhalos appears to be relatively independent of the mass of the
parent halo. Thus, trends such as those inferred by \scite{trenhodg}
would require processes that either preferentially suppress the
formation of dwarf galaxies in low mass systems, or destroy them after
they form. An effective mechanism for suppressing the formation of
small galaxies is the reheating of the intergalactic medium (IGM)
caused by the reionization of the Universe at a redshift $z\simeq
6$. \scite{tully02} have argued that this process could introduce an
environmental dependence in the galaxy luminosity function on the
grounds that a higher fraction of the low-mass halos that formed
before reionization (when dwarf galaxy formation proceeded unimpeded
by photoionization suppression) ended up in clusters today than in
less massive systems.

The effect of reionization on the formation of galaxies has been the
subject of several recent studies
\cite{bullock00,benson02,somerville02}, aimed mostly at investigating the
discrepancy between the large number of subhalos found in N-body
simulations of galactic CDM halos and the small number of satellite
galaxies observed in the Local Group. In this paper, we employ the CDM
model of \scite{benson02} to calculate the luminosity function of
galaxy populations residing in dark matter halos of different mass. We
find that there are significant differences in these luminosity
functions and we then explore how they combine together to build up
the global luminosity function, with particular emphasis on the faint
end slope.  A partial study of luminosity functions in halos of
different mass using a semi-analytic model of galaxy formation was
carried out by \scite{diaferio99}.

To calculate galaxy luminosity functions in halos of different mass
correctly, it is important to include tidal effects on satellite
galaxies, a potential galaxy destruction mechanism. Our model treats
these effects in considerably more detail than previous models of
galaxy formation. We find that tidal effects are important in limiting
the formation of massive galaxies at the centre of rich clusters. 

In this paper, we compare the results of our calculations to the data
of \scite{trenhodg} and assess whether feedback from reionization is a
viable explanation of the trend claimed by these authors. The existing
dataset is small, but forthcoming results from the 2dF and Sloan
galaxy surveys will enable much more extensive comparisons with the
theory.

The remainder of this paper is arranged as follows. In
\S\ref{sec:model} we briefly outline our model of galaxy formation, in
\S\ref{sec:res} we present results for the environmental dependence of
the luminosity function and in \S\ref{sec:compobs} we compare our
model with the available observational data. Finally, in
\S\ref{sec:dis} we present our conclusions. We present, in an
Appendix, several simple models of photoionization suppression to
elucidate how this mechanism works.

\section{Model}
\label{sec:model}

We employ the semi-analytic model of galaxy formation described in
detail by \scite{cole2000} and Benson et al.~(2002a; hereafter
Paper~I) to compute the properties of galaxies in a range of
environments at $z=0$. The reader is referred to those papers for a
complete and detailed description of the model. Briefly, the
hierarchical formation of dark matter halos is calculated using the
extended Press-Schechter formalism \cite{PS74,Bower91,Bond91}. The
formation of galaxies in the resulting dark matter halo merger trees
is followed by means of simple, physically motivated models of gas
cooling, star formation and galaxy merging. Recent work has
demonstrated that at least in so far as gas cooling is concerned these
simplified calculations are in excellent agreement with the results of
N-body/hydrodynamical simulations
\cite{ajbsph,yoshida02,helly02}. Applying a stellar population
synthesis model gives galaxy luminosities in different passbands.  The
model includes a prescription for supernovae feedback which drives gas
out of galaxies at a rate proportional to the current star formation
rate, with a constant of proportionality that is larger for less
strongly bound systems. This negative feedback flattens the faint end
slope of the luminosity function but, on its own, its effect is not
strong enough to account for the measured function
(Paper~I). \scite{benson02} developed, and incorporated into this
model, a detailed, self-consistent treatment of photoionization
suppression of galaxy formation (hereafter abbreviated as \PS). By
causing further suppression of faint galaxy formation, this mechanism
brought the model luminosity function into excellent agreement with
observations.  In Paper~I, we also included a much more detailed
treatment than was previously possible of the evolution of satellite
galaxies as they orbit within larger halos experiencing tidal forces
and gravitational shock heating. (Hereafter, we abbreviate ``tidal
limitation as \TL.)  The associated mass loss also affects the shape
of the luminosity function, and so we will explore this process in
this paper.

We adopt essentially the same model parameters as in Paper~I, with the
following differences. While the models of Paper~I used the
Press-Schechter halo mass function, we choose to adopt the function
proposed by \scite{jenkins01} which gives a better match to the
results of N-body simulations. Consequently, to produce a reasonable
galaxy luminosity function, we find it necessary to adjust the
parameters $\Omega_{\rm b}$ from $0.020$ to $0.024$, $\epsilon_\star$
(which determines the star formation rate) from 0.0050 to 0.0067, and
$\Upsilon$ (which affects mass-to-light ratios) from 1.38 to 1.03
(since $\Upsilon$ affects the fraction of mass recycled in star
formation, we adjust that fraction accordingly). Even with these
changes, the \scite{jenkins01} mass function results in a somewhat
worse fit to the bright end of the luminosity function than was
obtained in Paper~I. We defer further study of this aspect of the
model to future work. We adopt an escaping fraction of ionizing
photons from galaxies, $f_{\rm esc}$, of 12\%. This produces a neutral
hydrogen fraction for gas at zero overdensity in the IGM at $z=6$ of
$5\times 10^{-4}$, in agreement with the lower limit of $3.4\times
10^{-4}$ derived by \scite{lidz02}, while causing reionization to
occur at as high a redshift as possible. This maximizes the associated
suppression of galaxy formation\footnote{We could increase $f_{\rm
esc}$ slightly further without violating the limit of
\protect\scite{lidz02} but, since the neutral hydrogen fraction drops
very rapidly in our model at these redshifts (Paper~I), it makes
almost no difference to the suppression of galaxy formation.}. The
resulting model is essentially the same as the $f_{\rm esc}=10\%$
model presented in Paper~I.

\section{The Galaxy Luminosity Function}
\label{sec:LF}
\label{sec:res}

\subsection{Luminosity functions in halos of different mass}
\label{sec:haloLF}

We begin by studying the luminosity functions of galaxies in
individual dark matter halos of different mass, and then proceed to
consider how these combine together to form the global galaxy
luminosity function.  Using the model described in \S\ref{sec:model},
we simulate galaxy formation in a large number of halos of masses
$10^{11}$, $10^{12}$, $10^{13}$, $10^{14}$ and
$10^{15}h^{-1}M_\odot$\footnote{Throughout this paper we write
Hubble's constant as $H_0=100 h $km s$^{-1}$ Mpc$^{-1}$.}. This spans
halo masses appropriate to systems ranging from the Local Group to
rich galaxy clusters. (The $10^{11}h^{-1}M_\odot$ bin corresponds to
halos less massive than that of the Local Group but we include it
since it shows the most dramatic effects.) Merger trees are
constructed with a resolution of $10^{-5}$ of the final halo mass or
$10^8h^{-1}M_\odot$, whichever is smaller. This is sufficient to
follow accurately the formation and evolution of galaxies brighter
than $M_{\rm B}-5\log h = -10$ except in the most massive clusters
($10^{15}h^{-1}M_\odot$) for which resolution begins to affect our
results faintwards of $M_{\rm B}-5\log h = -12$; for these clusters,
we show results only down to $M_{\rm B}-5\log h = -13$. We perform
calculations for our ``standard model'' (i.e. including the effects of
\PS\ and \TL\ of satellite galaxies), and repeat them switching off
\PS\ and/or \TL\ (resulting in a total of four different sets of
results), in order to assess the effects of each physical mechanism
separately.  We simulate a sufficiently large number of halos of each
mass so that the average luminosity functions are reasonably smooth.

\begin{figure*}
\psfig{file=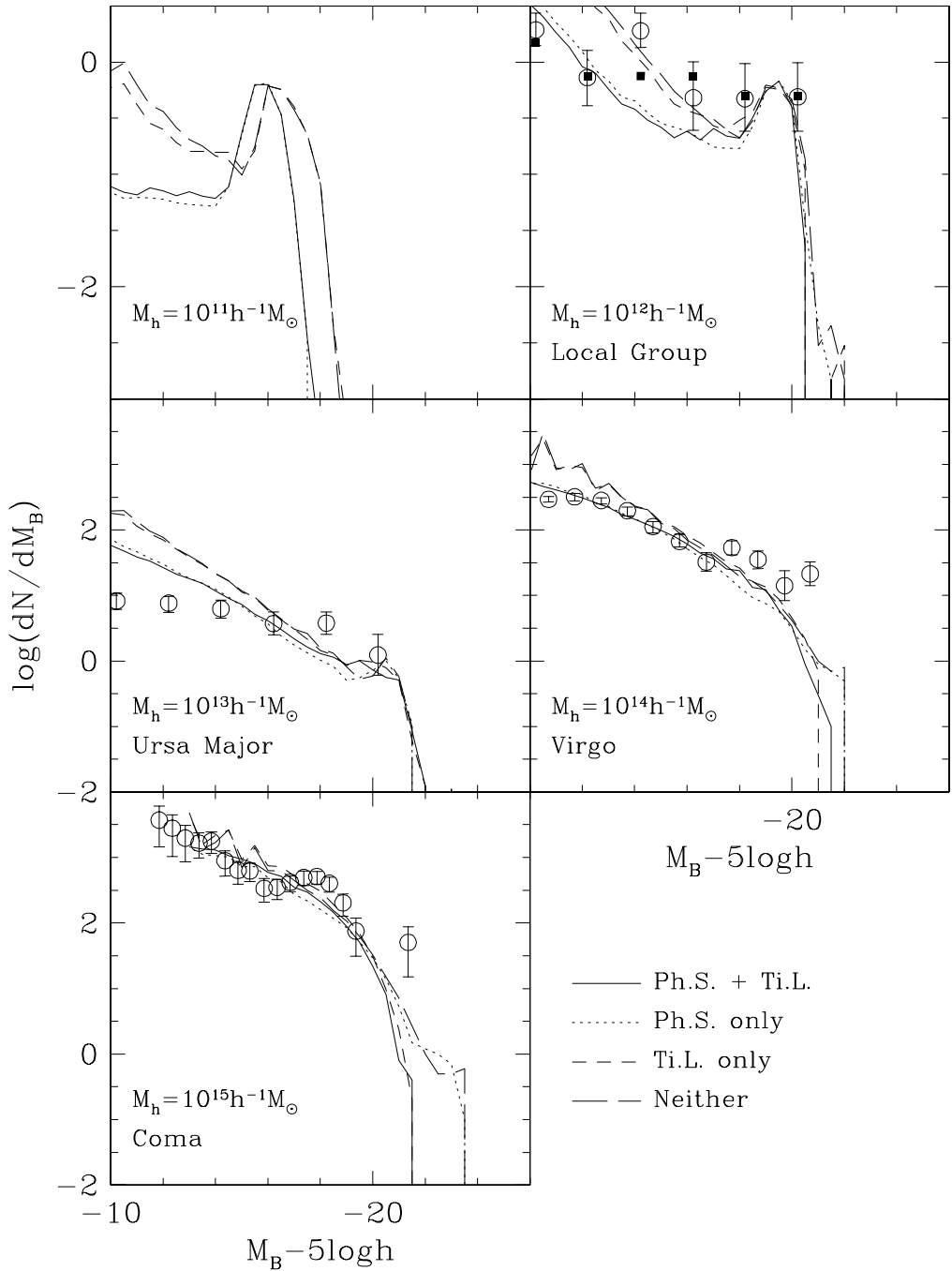,width=150mm}
\caption{B-band galaxy luminosity functions in halos of different
mass. Each panel shows the mean predicted model luminosity function in
an ensemble of dark matter halos of mass as given in each panel
(ranging from small galaxies to clusters). Specifically, we show the
mean number of galaxies per magnitude, per halo. Solid lines include
photoionization suppression (\PS) and tidal limitation (\TL), dotted
lines have \PS\ only, short-dashed lines have \TL\ only, and
long-dashed lines have neither. Open circles show observed luminosity
functions from the compilation of \protect\scite{trenhodg}, as
indicated in the legend of each panel. These have been normalized
arbitrarily to permit easier comparison of their shapes with the
models. Filled squares in the $10^{12}h^{-1}M_\odot$ panel show the
luminosity function compiled by \protect\scite{benson02b}, which
includes only galaxies classed as lying within the virial radii of the
Milky Way's or M31's dark halos. The comparison between the model and
the observational data is discussed in \S\ref{sec:compobs} }
\label{fig:eLF}
\end{figure*}

The solid lines in Figure~\ref{fig:eLF} show the luminosity functions
of galaxies in our simulated halos.  The basic form of the standard
model luminosity function is similar in all mass halos. It has two
components: a near power-law distribution at faint magnitudes which is
made up of satellite galaxies (i.e. those galaxies that do not reside
at the centre of the final halo), and a strongly peaked distribution
at bright magnitudes beyond which no more galaxies are found and which
is made up of the central galaxy in each final halo. As we consider
halos of increasing mass, both components shift to ever brighter
luminosities, but the peaked component decreases in amplitude relative
to the power-law component, indicating that, in low mass halos, a
single galaxy tends to dominate, while in clusters the light is more
equally shared among many galaxies.

Observed luminosity functions, $\phi(L)=\d n/\d L$ (where $n$ is the
number of galaxies per unit volume and $L$ is luminosity), are
typically described by the Schechter function,
\begin{equation}
\phi(L) \d L = \phi_* \left( {L \over L_*} \right)^\alpha \exp \left( - {L \over L_*} \right) {\d L \over L_*},
\end{equation}
where $\phi_*$, $L_*$ and $\alpha$ are parameters. At magnitudes faint
compared to the characteristic luminosity, $L_*$, has the simple form
$\phi(L) \propto L^\alpha$.  Following normal practice, we will refer
to $\alpha$ as ``the faint end slope.''  Although the luminosity
functions in Fig.~\ref{fig:eLF} are not well fit by the Schechter
form, we can still obtain a formal value of $\alpha$ by fitting to the
data in the first few bins plotted. We find slopes in the range
$-1.57<\alpha<-1.48$, except in the lowest mass halos
($10^{11}h^{-1}M_\odot$) for which a much flatter slope,
$\alpha=-1.06$, is found (although it is possible that a steeper slope
would be obtained also in this case if we probed to even fainter
luminosities). There is little evidence for a strong variation of
slope with halo mass, except perhaps for a weak (and rather
insignificant) trend for \emph{flatter} slopes in higher mass
clusters.  However, it is important to note that the slope is not
constant with galaxy magnitude. This is clearly seen, for example, in
the $10^{12}h^{-1}M_\odot$ halos for which, in the region $-18<M_{\rm
B}-5\log h<-15$, the slope is close to $\alpha=-1$.

\begin{figure*}
\begin{tabular}{cc}
\psfig{file=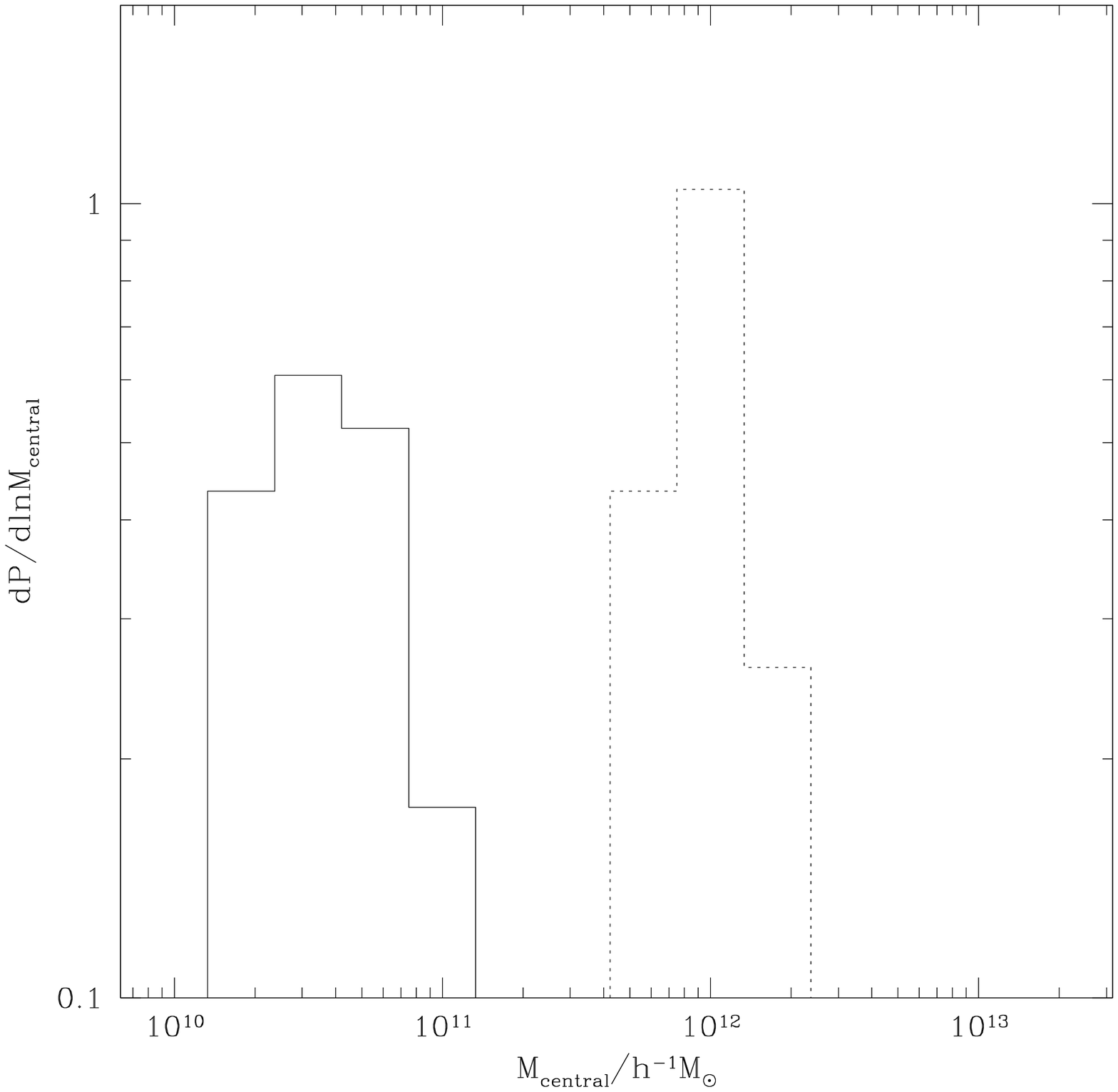,width=80mm} & \psfig{file=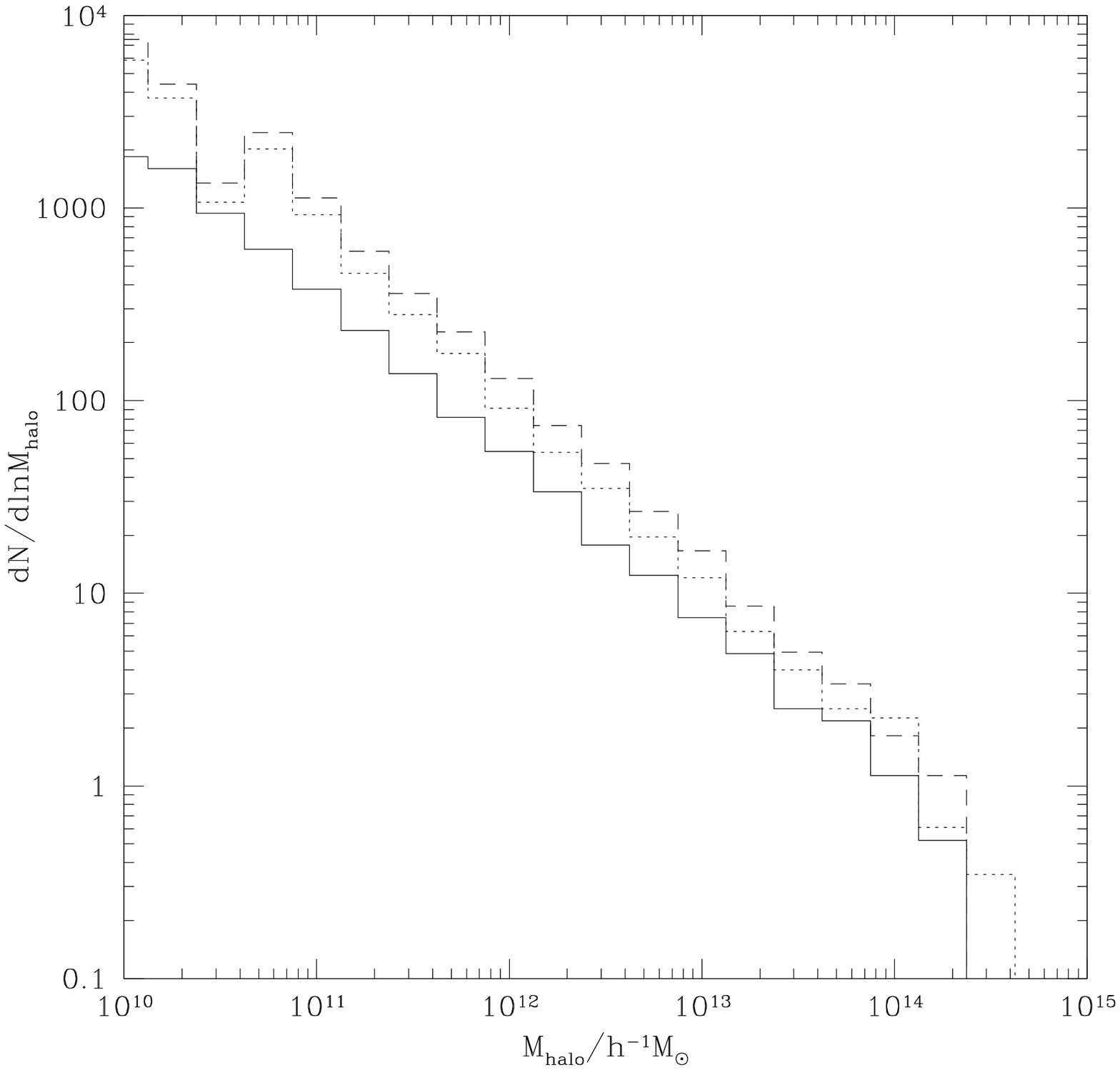,width=80mm}
\end{tabular}
\caption{\emph{Left-hand panel:} The distribution of baryonic masses
of galaxies residing at the centres of $10^{15}h^{-1}M_\odot$ clusters
at $z=0$. The solid histogram shows results from our standard model
which includes \TL\ while the dotted histogram corresponds to a model
with no \TL. \emph{Right-hand panel:} The mass function of
\emph{surviving} substructure halos in $10^{15}h^{-1}M_\odot$ clusters
at $z=0$. The dotted histogram is from a model with no \TL, while
dashed and solid histograms are from the model with \TL\ and give 
the masses of halos before and after tidal mass loss respectively.}
\label{fig:tidal}
\end{figure*}

Switching off \TL\ in our model, and reverting to the simpler
calculation of dynamical friction of \scite{cole2000}, results in the
dotted curves of Fig.~\ref{fig:eLF}. Clearly, \TL\ has little
influence on the faint end of the luminosity function, although it
does slightly reduce the luminosities of already faint galaxies (as is
particularly noticeable for $10^{12}h^{-1}M_\odot$ halos). A more
interesting effect of \TL\ is apparent in the massive clusters
($10^{14}$ and $10^{15}h^{-1}M_\odot$ halos). Here, switching off \TL\
results in a new population of very luminous galaxies, visible as a
``bump'' in the luminosity function at bright magnitudes. These highly
luminous objects form through multiple mergers at the centres of
cluster halos. When \TL\ is included, the merger rate of massive
galaxies in clusters is significantly reduced, and none of these
highly luminous objects form. As shown in the left-hand panel of
Fig.~\ref{fig:tidal}, cluster central galaxies are about an order of
magnitude less massive when our accurate calculation of dynamical
friction is used. Note that the inclusion of \TL\ makes much less
difference in $10^{12}h^{-1}M_\odot$ halos, due to the very different
shape of the luminosity function in these systems (i.e. there is
relatively little mass in satellites to merge into the central
object). This difference between the merger rates in our model and
that of \scite{cole2000} cannot be fully reproduced by simply
lengthening the dynamical friction timescale in the \scite{cole2000}
prescription by some factor (even if this factor is a function of halo
mass). The shape of the distribution of merger times is significantly
different between the models and, in our calculation, varies with halo
mass.

An effect similar to that shown by the comparison of the solid and
dotted lines in Fig.~\ref{fig:eLF} was noticed by \scite{springel01}
in cluster luminosity functions calculated using a different
semi-analytic model. \scite{springel01} computed luminosity functions
adopting both simple estimates of galaxy merger rates (similar to
those assumed in computing the dotted curves in Fig.~\ref{fig:eLF})
and merger rates taken directly from a high resolution simulation of
dark matter. They found that the simple scheme produced
highly luminous galaxies through merging that were not present in the
N-body calculation. Based on the number of massive mergers quoted by
\scite{springel01}, we can conclude that a similar reduction in
central galaxy mass occurred when our realistic merger rate
calculation was employed.  As suggested by
\scite{springel01}, the cause of this difference is that the simple
scheme underestimates the merging timescale of rather massive
subhalos in clusters, leading to many massive galaxies merging with
the central object in the final halo. In our more detailed
calculation, as in the N-body simulation, these large subhalos loose
a significant amount of their mass as they orbit within the cluster,
leading to a reduced dynamical friction force and hence to a reduced
merger rate.

The right-hand panel of Fig.~\ref{fig:tidal} shows the mass functions
of \emph{surviving} substructure halos in clusters at $z=0$. The
dotted histogram shows the model with no \TL, while the dashed
histogram corresponds to the model with \TL\ included, but the mass of
each substructure is plotted as it was \emph{before} it experienced
any tidal mass loss. Clearly, inclusion of \TL\ has resulted in fewer
mergers since there are more surviving substructures of a given
initial mass. The solid histogram gives the distribution of
substructure masses after tidal mass loss. Comparing the dashed and
solid histograms indicates that halos typically lose between 50\% to
70\% of their mass (with higher mass halos losing less than lower mass
halos). The dynamical friction time scales as the mass of the
satellite halo and is therefore increased by factors of 2 to 3 when
\TL\ is included, greatly reducing the galaxy merger rates.

Switching off \PS\ (short-dashed lines in Fig.~\ref{fig:eLF}) has a
more dramatic effect on the luminosity functions than switching off
\TL.  Firstly, the faint end slope steepens in almost all systems to
values in the small range $-1.71<\alpha<-1.64$. (The exception are the
lowest mass halos which have $\alpha=-1.38$ because of the
particularly strong effects of supernovae feedback in these halos.)
The increase in the number of faint galaxies is more pronounced in
lower mass systems. In high-mass systems there is a weak trend for the
bright end of the luminosity function to be shifted slightly
faintwards when \PS\ is ignored. In this case, more gas is locked up
in the fainter galaxies, leaving less gas to form massive, bright
galaxies. In the lowest mass systems the opposite effect occurs and
the central galaxies become brighter when \PS\ is ignored. In these
$10^{11}h^{-1}M_\odot$ halos, \PS\ suppresses the amount of gas that
is able to accrete into the halo, thus reducing the luminosity of the
associated galaxy. The way in which various processes associated with
photoionisation shape the luminosity function in halos of different
mass, particularly its faint end, is discussed in detail in the
Appendix. There we also discuss the effect of various approximations
in the treatment of photoionisation.

Long-dashed lines show results when both \TL\ and \PS\ are left out of
our calculations. These are essentially accumulations of the effects
seen when each process was switched off individually, and result in a
further steepening of the faint end slopes. In these models, only
feedback from supernovae affects the faint end of the luminosity
function in a manner which is essentially independent of halo mass. 

\subsection{The global luminosity function} 
\label{sec:globalLF}

We now explore how the individual halo luminosity functions presented
in Section~\ref{sec:haloLF} combine to produce the global
luminosity function. An initially surprising aspect of the individual
halo luminosity functions is that their faint end slopes are, in all
cases, steeper than that of the global luminosity function. This, as
we showed in Paper~I, has a slope of around $-1.2$ both in our models
and in the real universe. How does \PS\ contrive to produce a flatter
faint end slope for the total population than for any of the
constituent individual halos?  To understand this behaviour, we 
must consider the contribution of different mass halos to the global
luminosity function. 

\begin{figure*}
\psfig{file=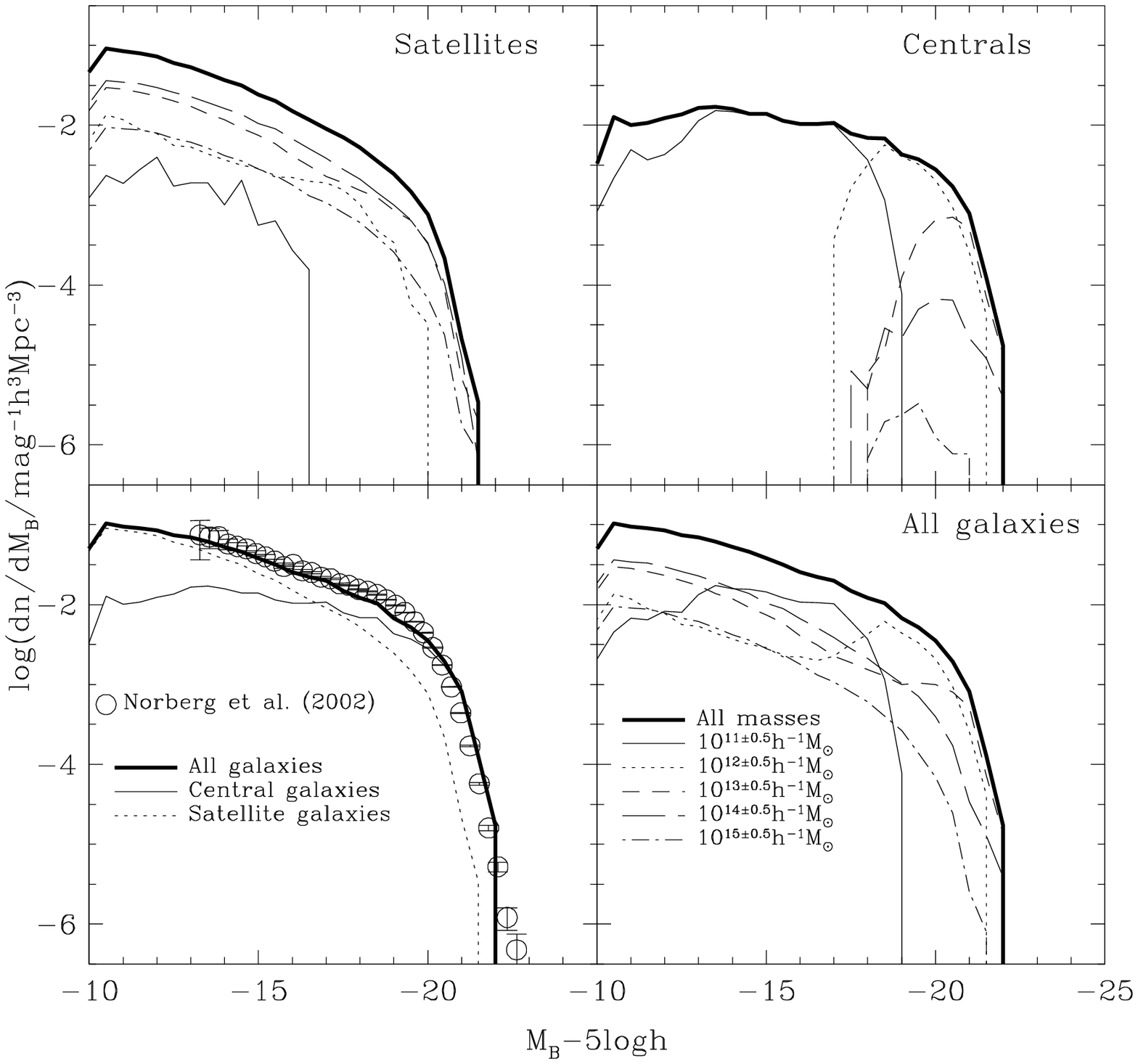,width=160mm,bbllx=10mm,bblly=90mm,bburx=190mm,bbury=265mm,clip=}
\caption{B-band galaxy luminosity functions.  \emph{Upper left-hand
panel:} The luminosity functions of satellite galaxies (i.e. those
which are not the central galaxy of their host halo). The heavy solid
line shows the luminosity functions summed over all halo masses, while
thin lines show the contributions from halos in different mass ranges
as indicated by the key in the lower right panel. \emph{Upper
right-hand panel:} The luminosity functions of central galaxies. Line
types are as in the upper left-hand panel. \emph{Lower left-hand
panel:} A comparison of the satellite (dotted line) and central galaxy
(thin solid line) luminosity functions. Also show is the total
luminosity function (heavy solid line) and the observational
determination of the total luminosity function from
\protect\scite{norberg02}. The contributions from all types of galaxy
(i.e. satellites and centrals) in halos of different mass to the total
luminosirty function. Line types are as in the upper left-hand panel.}
\label{fig:LFcont}
\end{figure*}

The upper panels of Fig.~\ref{fig:LFcont} show the contributions to
the global luminosity function from satellite and central galaxies
(left and right-hand panels respectively) residing in halos of
different mass. The slope of the combined satellite luminosity
function reflects the power-law slope of the individual halo
luminosity functions and is dominated over a wide range of
luminosities by halos of $10^{13}$--$10^{14}h^{-1}M_\odot$. The
luminosity function of central galaxies, on the other hand, has a much
flatter faint end slope, which is determined by the way in which
central galaxy luminosity scales with halo mass. In our models, this
scaling depends primarily on the combined effects of supernovae and
\PS\ feedback. Since the central galaxy luminosity functions of
individual halos are strongly peaked around a particular luminosity,
the global central galaxy luminosity function is dominated by halos in
a narrow range of mass at each luminosity. (The peak for galaxies in
$10^{10.5}$ to $10^{11.5}h^{-1}M_\odot$ is much broader, since these
mass scales are affected by \PS, causing strong suppression of galaxy
formation in the lower mass halos in the range. Figure~\ref{fig:eLF}
shows much more clearly the sharply peaked luminosity distribution of
central galaxies in halos of fixed mass.)

The lower left-hand panel of Fig.~\ref{fig:LFcont} shows how the
satellite and central galaxy luminosity functions combine to produce
the global luminosity function (which is compared to the recent
observational determination from the 2dF galaxy redshift survey of
\pcite{norberg02}). Brightwards of $M_{\rm B}-5\log h\approx -17$,
central galaxies dominate, while faintwards of this, the satellite
contribution gradually takes over. The faint end slope of the global
luminosity function ends up being intermediate between that of the
satellite and central galaxy luminosity functions.

Finally, in the lower right-hand panel of Fig.~\ref{fig:LFcont} we
show the contribution to the global luminosity function from halos of
different mass. Clearly, the faint end slope of the global luminosity
function is determined in part by the relation between central galaxy
luminosity and halo mass, not just by the faint end slopes of the
individual halo luminosity functions. This explains why the global
galaxy luminosity function ends up having a flatter faint end than the
individual halo luminosity functions. It is interesting to note that
both the total and cluster (i.e. $10^{15}h^{-1}M_\odot$ halo)
luminosity functions are reasonably well fit by Schechter functions,
albeit with slightly different parameters ($\alpha=-1.31/-1.45$ and
$M_*-5\log h=-20.12/-19.60$ for the total/cluster luminosity function;
note that the correlation between the parameters $\alpha$ and $M_*$
should be borne in mind when comparing these values).

\section{Comparison with Observations}
\label{sec:compobs}

Our model predicts marked differences in the luminosity functions of
galaxy populations residing in halos of different mass. Although the
detailed form of these predictions depends on the details of our
galaxy formation model, the gross differences seen in
Fig.~\ref{fig:eLF} are generic predictions that bear directly on basic
features of the model such as hierarchical clustering from CDM initial
conditions, gas cooling, photoionization, feedback, etc. Thus, in
principle, individual halo luminosity functions provide a strong test
of the CDM galaxy formation paradigm. Unfortunately, implementing such
a test in practice is difficult because of the observational
complications inherent in identifying galaxies attached to a
particular dark matter halo. Gravitational lensing is a powerful and
promising tool for detecting dark matter halos directly, but this
technique is still in early stages of development (e.g.
\pcite{Mellier02}). A less direct, but still useful approach, consists
of finding groups and clusters in large redshift catalogues such as
the 2dF and SDSS surveys. When interpreted with the aid of
cosmological simulations, it is possible to go some way towards
identifying galaxy populations likely to be associated with single
halos (and their subhalos) of a given mass \cite{Eke02}. This
approach will yield interesting data for comparison with our model
predictions in the near future.

In the interim, \scite{trenhodg} have provided a limited compilation
of observational data. They give estimates of the B-band luminosity
functions of galaxies in the Local Group and in the Ursa Major, Virgo
and Coma clusters. In our model, these systems are expected to reside
in halos of mass similar to those we have simulated, i.e. $10^{12}$,
$10^{13}$, $10^{14}$ and $10^{15}h^{-1}M_\odot$ respectively
\cite{benson02,trenhodg,schindler99,geller99}. We show Trentham \&
Hodgkin's (2002) compilation in Fig.~\ref{fig:eLF} as open circles.
The vertical normalization of the datasets have been adjusted
arbitrarily to permit an easier comparison of the shapes and faint end
slopes of the model and observed luminosity functions. For the two
most massive systems in the figure, our standard model is in
reasonably good agreement with the observed luminosity function,
including the faint end slope, but for Ursa Major and the Local Group,
the observations indicate a much shallower slope than is produced by
our model. The same discrepancy was noted by \scite{benson02b} for the
Local Group, and is also apparent in the predicted Local Group
luminosity function of \scite{somerville02} (which has
$\alpha\approx-1.5$ over the range $-20<M_{\rm V}-5\log h_{70}<-10$,
where $h_{70}$ is the Hubble constant in units of 70km/s/Mpc, for the
case of reionization at $z=8$ which is the closest to our own
calculations), suggesting that this is a rather robust prediction of
\PS.

In making the comparison between theory and observations in
Fig.~\ref{fig:eLF}, one should keep in mind the possibility that the
two may not correspond to quite the same quantity. The theoretical
predictions pertain to luminosity functions of galaxy populations
residing in individual dark matter halos of a given mass. However,
there is no guarantee that the samples selected by \scite{trenhodg} do
indeed come from individual dark matter halos.  For example, these
samples might be contaminated by contributions from secondary halos
which happen to be nearby the dominant halo. This could be
particularly important for the Ursa Major cluster which is a rather
diffuse object and is too large, given its mass, to be associated with
a single, virialized halo (according to the standard theoretical
definitions).  There are ambiguities regarding the Local Group data as
well. As argued by \scite{benson02b}, many dwarfs nominally associated
with the Local Group lie beyond the region which, according to our
model, contains the (distinct, but similar mass) halos of the Milky
Way and M31. In Fig.~\ref{fig:eLF}, we show as filled squares the
luminosity function of the galaxies that \scite{benson02b} would class
as lying within the virial radius of the Milky Way or M31.  This
luminosity function also shows a fairly flat faint end. Nevertheless,
selection effects such as these must be accounted for in a more
thorough comparison of theory and observations.

\begin{figure*}
\psfig{file=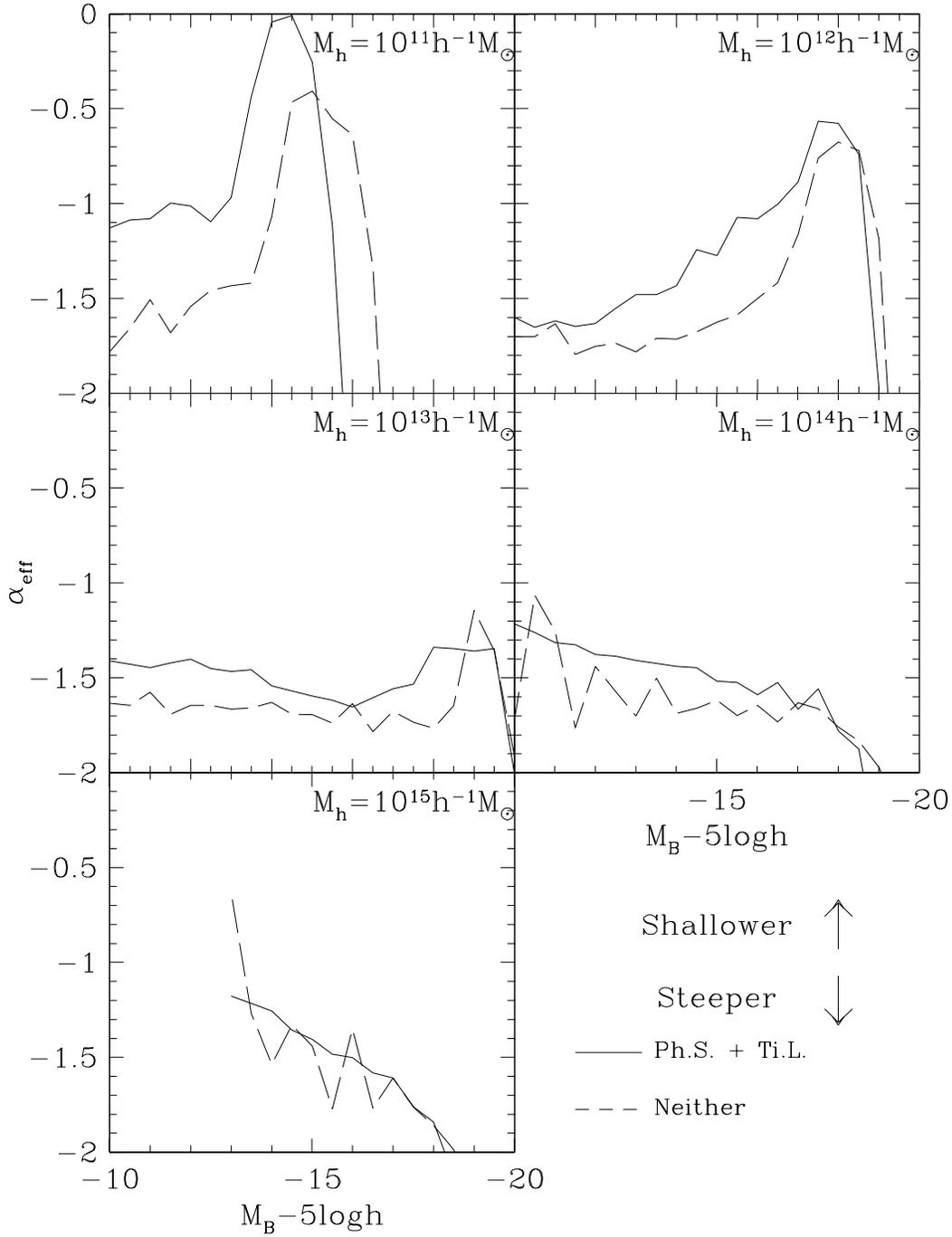,width=150mm}
\caption{Effective slope of model luminosity functions 
{\it vs.} absolute magnitude for galaxy populations residing in halos
of different mass, as indicated in the panels. Solid lines correspond
to the model that includes both \PS\ and \TL, while the dashed lines
correspond to the model that includes neither.}
\label{fig:ealpha}
\end{figure*}

The results of Fig.~\ref{fig:eLF} are shown in a different way in
Fig.~\ref{fig:ealpha} which more clearly shows the variation in the
shapes of the luminosity functions at different luminosities.  Here,
we plot the effective luminosity function slope, $\alpha_{\rm eff}= L
(\d^2 N/\d L^2)/(\d N/\d L)$, as a function of absolute magnitude for
all the systems we have simulated. (For clarity, we show results only
for our standard model and for a model with both \PS\ and \TL\
switched off.)  The luminosity functions with no \PS\ or \TL\ in
$10^{13}$ and $10^{14}h^{-1}M_\odot$ halos show an extended region
where $\alpha_{\rm eff}$ is almost constant, indicating a power-law
luminosity function. With \PS\ and \TL\ switched on, the slope is
shallower, but the luminosity function is no longer a power-law over
an appreciable range of magnitudes. For the other mass systems, the
luminosity function is nowhere well described by a power-law and, in
every case, the inclusion of \PS\ makes the slope shallower, the
effect being larger in the lower mass systems.

\scite{trenhodg} characterized variations in luminosity function
shapes using the ``dwarf-to-giant ratio'' (i.e. the number of galaxies
in some range of faint magnitudes relative to the number at brighter
magnitudes). We compute a similar ratio for our model with \PS\ and
\TL\ and present the results in Table~\ref{tb:dgrat}.
(We use a slightly different magnitude range for the dwarfs because,
for our most massive halos, our model is not complete over the range
of magnitudes used by Trentham \& Hodgkin.)

\begin{table*}
\caption{Dwarf-to-giant ratios for galaxy populations in different
environments. The magnitude ranges for dwarfs and giants are specified
in terms of $M_{\rm B}-5\log h$ (to maintain consistency with the rest
of this work), and correspond to $-16<M_{\rm B}\leq-14$ and $M_{\rm
B}\leq-16$ respectively for $h=0.7$.}
\label{tb:dgrat}
\begin{tabular}{ccc}
\hline
Halo mass ($h^{-1}M_\odot$) & $N(-15.23<M_{\rm B}-5\log
h\leq-13.23)/N(M_{\rm B}-5\log h\leq-15.23)$ & Observed \\ 
\hline
$10^{11}$ & 0.24 & --- \\
$10^{12}$ & 0.40 & $1.25 \pm 0.35$ \\
$10^{13}$ & 1.85 & $0.74 \pm 0.07$ \\
$10^{14}$ & 1.70 & $1.47 \pm 0.18$ \\
$10^{15}$ & 1.68 & $1.40 \pm 0.11$ \\
\hline
\end{tabular}
\end{table*}

For the two most massive systems, our model results are in reasonably
good accord with the observational determinations, but for the lower
mass systems there are significant discrepancies. This analysis
verifies and quantifies the comparison between the shapes of the
predicted and observed luminosity functions made in
Fig.~\ref{fig:eLF}.

\section{Discussion}
\label{sec:dis}

Using a semi-analytic model of galaxy formation, we have calculated
the luminosity function of galaxy populations contained in halos of
different mass. We find large differences in the shapes of these
luminosity functions. In smaller mass halos ($M\lsim 10^{13}
h^{-1}M_\odot)$, the luminosity function has a ``hump'' at bright
magnitudes and a roughly power-law shape at fainter magnitudes. In
more massive systems, the shape approaches the familiar Schechter
form. These differences reflect the relative contributions of
``central'' and ``satellite'' galaxies. In the smaller mass halos, the
bright end of the luminosity function is dominated by a single or at
most a few bright galaxies and the rest of the population consists
predominantly of much fainter satellites. In the more massive halos,
the light is more evenly distributed amongst a large number of
galaxies. These differences reflect the complex interplay between the
processes that establish the number of halos and subhalos of different
mass (i.e. hierarchical clustering from CDM initial conditions) and
the processes that light them up (i.e. gas cooling, stellar evolution,
feedback, etc).

The global luminosity function is the weighted sum of the luminosity
functions of individual halos. Interestingly, the faint end slopes of
the individual luminosity functions are always steeper than the
faint end slope of the global luminosity function.  This can be
understood in terms of the relative contributions to the global
luminosity function of satellite and central galaxies from individual
halos, whose luminosity functions have steep and shallow slopes
respectively. The slope of the global function is intermediate between
these two.

Our model of galaxy formation represents an advance over previous
models because of its detailed treatment of two processes that are
important in establishing the luminosity function: the suppression of
galaxy formation in small halos at high redshift (\PS), as a result of
the photoionization of the IGM by early generations of galaxies and
quasars, and tidal mass loss from satellite galaxies (\TL). \PS\ is
particularly relevant at the faint end and has the net effect of
reducing the number of faint galaxies, giving rise to a flatter
luminosity function than would otherwise be the case. This outcome is
far from obvious. For example, a simplistic model of \PS\ in which
galaxy formation is assumed not to occur in low mass halos below the
reionization redshift results, in fact, in a steeper luminosity
function (see Appendix~\ref{sec:faint}). In reality, the effects of
\PS\ depend on the mass of the halo relative to the time-dependent
``filtering mass'' (defined as the halo mass which accretes only half
the mass that it would have accreted in the absence of
photoionization; \pcite{gnedin00}). Galaxies can form even if their
halos have mass smaller than the filtering mass, but with reduced
luminosity.  The lower the mass of the halo relative to the filtering
mass, the fainter is the galaxy that forms. The filtering mass is an
increasing function of time. Thus, although lower mass halos form at
higher redshift, the redshift where the filtering mass becomes
comparable to their mass is also higher, increasing the fraction of
them which experience strong suppression. These two effects combined
\emph{do} tend to produce a flatter luminosity function but the degree
of flattening depends on their detailed balance\footnote{A model in
which a cut-off in circular velocity rather than mass is assumed gives
results in quite good agreement with our detailed calculations
\cite{benson02}.}. As we have shown previously (\pcite{benson02}; see
also \pcite{benson02b}), \PS\ on its own cannot account for the
shallow faint end slope of the global luminosity function. Other
processes, such as feedback from energy released by stellar winds and
supernovae, need to be included in order to reduce the number of faint
galaxies to the observed level.

The other process that we have treated in detail, \TL, is relevant at
the bright end of the luminosity function, particularly in massive
halos harbouring rich clusters. Mass loss from satellite galaxies, as
they spiral into the centre of their halo, increases the dynamical
friction time, so reducing the merger rate. This prevents the
formation of some of the highly luminous and unobserved cluster
galaxies that tend to appear in models of galaxy formation that do not
take this process into account.

Keeping in mind the provisos of \S\ref{sec:compobs}, we find that our
model fails to reproduce the steepening of the faint end slope of the
luminosity function with system richness found by \cite{trenhodg}. The
model agrees quite well with the data for the Virgo and Coma clusters,
but it predicts a steeper faint end than is inferred for the Local
Group and Ursa Minor. Photoionization is an effective (and
unavoidable) mechanism for suppressing galaxy formation in low mass
halos, but it does not produce the kind of ``environmental'' variation
seen by \scite{trenhodg}. At the low mass end, this failure of the
model is related to the discrepancy that we (\pcite{benson02b}, see
also \pcite{somerville02}), found previously in our CDM models of the
Local Group. Photoionization has a dramatic effect on the overall
number of satellites that survive in galactic halos and can reduce it
to the levels seen in the Local Group. However, the satellite
luminosity function in the models is considerably steeper than is
observed. This discrepancy contrasts with the success of CDM models in
reproducing the structural properties of the satellites
\cite{stoehr02}.

Taking at the face value the discrepancy between our models and the
data of \scite{trenhodg}, it is interesting to speculate about the
possible root causes of the problem. Leaving aside alternative
theories for the nature of the dark matter or drastic changes to the
initial power spectrum of density perturbations specifically designed
to reduce the amount of small-scale power (e.g.
\pcite{colin00,kamion00}), we focus on our modeling of baryonic
processes within the context of the CDM cosmogony. This is undoubtedly
simplified. For example, the effects of \PS\ depend upon both the time
variation of the filtering mass and on the assumed functional form of
the suppression, both of which are uncertain. As discussed in the
Appendix, reasonable variations in our assumed mass dependence of \PS\
do not seem capable of producing a sufficiently flat faint end slope
in our models of the Local Group. However, we cannot exclude the
possibility that the simplifications made in our model are
inappropriate.  For example, we neglect the patchy nature of
reionization, but if low density regions became ionized first, galaxy
formation would be suppressed more effectively in these regions,
altering the temporal evolution of the filtering mass.  Localized \PS\
triggered, for example, by a nearby galaxy or quasar could also change
our model predictions. These processes are poorly understood
theoretically, but observational evidence (e.g. \pcite{kurt02}) 
suggests that they could be important in determining the thermodynamic
state of the IGM at high redshift. 

At a more basic level, there are approximations in our model which
might be inappropriate in the regime of low mass halos.  Thus,
although comparisons with N-body/hydrodynamical simulations indicate
that our estimates of gas cooling rates are accurate for relatively
massive galaxies \cite{ajbsph,yoshida02,helly02}, it has not been
shown that the same is also true for the low mass galaxies of interest
here. Cooling times in these objects can become very short and this
might conceivably render incorrect the assumption of cooling from an
initial, spherically symmetric quasi-hydrostatic equilibrium.  Further
study of gas cooling in low mass systems is necessary to address this
issue. Similarly, it is far from clear that the simple rules for star
formation in our model, which are empirical parameterizations tuned to
match observational constraints for relatively bright galaxies at low
redshifts, remain valid when extrapolated to low mass galaxies and/or
high redshifts. Other physical constraints on star formation not
currently included in our models may also be important for these low
mass galaxies (e.g. \pcite{lrp02}). Finally, our modeling of feedback
is also simplified and, again, its extrapolation to low masses may be
unrealistic. Possible sources of feedback such as strong outflows or
heating by AGN which are currently neglected may be important on these
scales.

In conclusion, we have found substantial differences in the luminosity
functions of galaxies residing in dark matter halos of different
mass. These result from the interplay of a variety of processes that
affect the formation of galaxies of different luminosity. Although the
exact predictions of our model no doubt depend on its details, the
gross differences between the predicted luminosity functions are
likely to be generic and to be present in a broad class of cold dark
matter models of galaxy formation. Comparison with existing data gives
mixed results: fair agreement with the luminosity functions in Virgo
and Coma but substantial disagreement with the luminosity functions in
the Local Group and Ursa Minor cluster. The much larger and better
controlled samples that will be forthcoming from cluster analyses of
the 2dF and SDSS galaxy redshift surveys will allow a more
comprehensive test of the general class of cold dark matter models of
galaxy formation.

\section*{Acknowledgments}

\appendix

\section{Photoionization and the Faint End of the Luminosity Function} 
\label{sec:faint}

The faint end of the luminosity function directly reflects the outcome
of feedback effects. The two sources of feedback in our model,
supernova-driven winds and photoionization, are both important in
determining the number of faint galaxies. As we discussed in Paper~I,
photoionization inhibits the formation of faint galaxies for two
reasons: it raises the temperature and pressure of the IGM, curtailing
the ability of gas to accrete into halos, and it reduces the rate at
which gas in halos can cool, inhibiting star formation. The first of
these processes is the dominant effect. It can be conveniently
described in terms of a ``filtering mass,'' $M_{\rm F}$, defined as
the mass of a halo which accretes only half the baryonic mass that it
would have accreted in the absence of photoionization. The mass of gas
accreted by a halo of mass $M_{\rm h}$ is approximated by the
following formula \cite{gnedin00}, based on the results of
gas-dynamical simulations:
\begin{equation}
M_{\rm gas} = {f_{\rm b} M_{\rm h} \over [1+(2^{1/3}-1)M_{\rm F}/M_{\rm h}]^3},
\end{equation}
where $f_{\rm b}$ is the universal baryon fraction and $M_{\rm F}$ is
the filtering mass.

There are three different aspects to consider when evaluating the
effect of the filtering mass on the faint end slope of the luminosity
function:

\begin{enumerate}
\item Galaxy formation is significantly suppressed in halos 
less massive than the filtering mass.
\item The filtering mass is an increasing function of time (at least
over the range of redshifts of interest here).
\item For halos formed at a given redshift, the suppression of galaxy
formation is greater the smaller the halo mass is relative to the
filtering mass. 
\end{enumerate}

The impact of these three factors will depend on the distribution of
formation redshifts of dark matter halos of different mass in
different environments\footnote{By ``environment'' we mean the mass of
the halos into which that earlier halo has been incorporated by the
present day.}. According to the extended Press-Schechter formalism, it
is always true (for a CDM power spectrum) that the typical formation
redshift of a halo increases as the mass of the halo decreases, and
also as the mass of the final host in which the halo resides
increases. Further understanding of halo formation redshift
distributions may be gained by considering eqn.~2.15 of \scite{lc93}
that gives the mass function of progenitor halos at any redshift and
which we reproduce in slightly different form below:
\begin{eqnarray}
{\d N \over \d M} (z) & = & {1 \over \sqrt{2\pi}} {M_{\rm host}\over M} {(\delta_{\rm c}^{\rm (z)}-\delta_{\rm c}^{\rm (0)})\over [\sigma^2(M)-\sigma^2(M_{\rm host})]^{3/2}} \nonumber \\
 & & \exp \left( -{(\delta_{\rm c}^{\rm (z)}-\delta_{\rm c}^{\rm (0)})^2\over 2[\sigma^2(M)-\sigma^2(M_{\rm host})]} \right) {\d \sigma^2 \over \d M},
\label{eq:PS}
\end{eqnarray}
where $\delta_{\rm c}^{\rm (z)}$ is the critical value of the linear
theory fractional overdensity for collapse at redshift $z$, $M$ is the
mass of the progenitor halo at that redshift, $\sigma^2(M)$ is the
variance of the linear fractional overdensity in a sphere of mass $M$,
and $M_{\rm host}$ is the mass of the halo into which that progenitor
has been incorporated by the present day.

Consider galaxies living in relatively low mass halos at the present
day (in which most of the galaxies that make up the faint end of the
luminosity function reside). The relative numbers of halos of masses
$M_1$ and $M_2(<M_1)$ at some particular redshift depends upon
$\sigma^2(M_1)-\sigma^2(M_{\rm host})$ and
$\sigma^2(M_2)-\sigma^2(M_{\rm host})$. As $M_{\rm host}$ increases,
$\sigma^2(M_{\rm host})$ decreases for a CDM power spectrum. Thus, for
sufficiently large $M_{\rm host}$, the relative numbers of halos of
masses $M_1$ and $M_2$ tends to a fixed value. For $M_{\rm host}$
comparable to $M_1$, the relative numbers of these halos become a very
strong function of $M_{\rm host}$, and the number of mass $M_1$ halos
is exponentially suppressed, as indicated by eqn.~(\ref{eq:PS}).

\begin{figure*}
\psfig{file=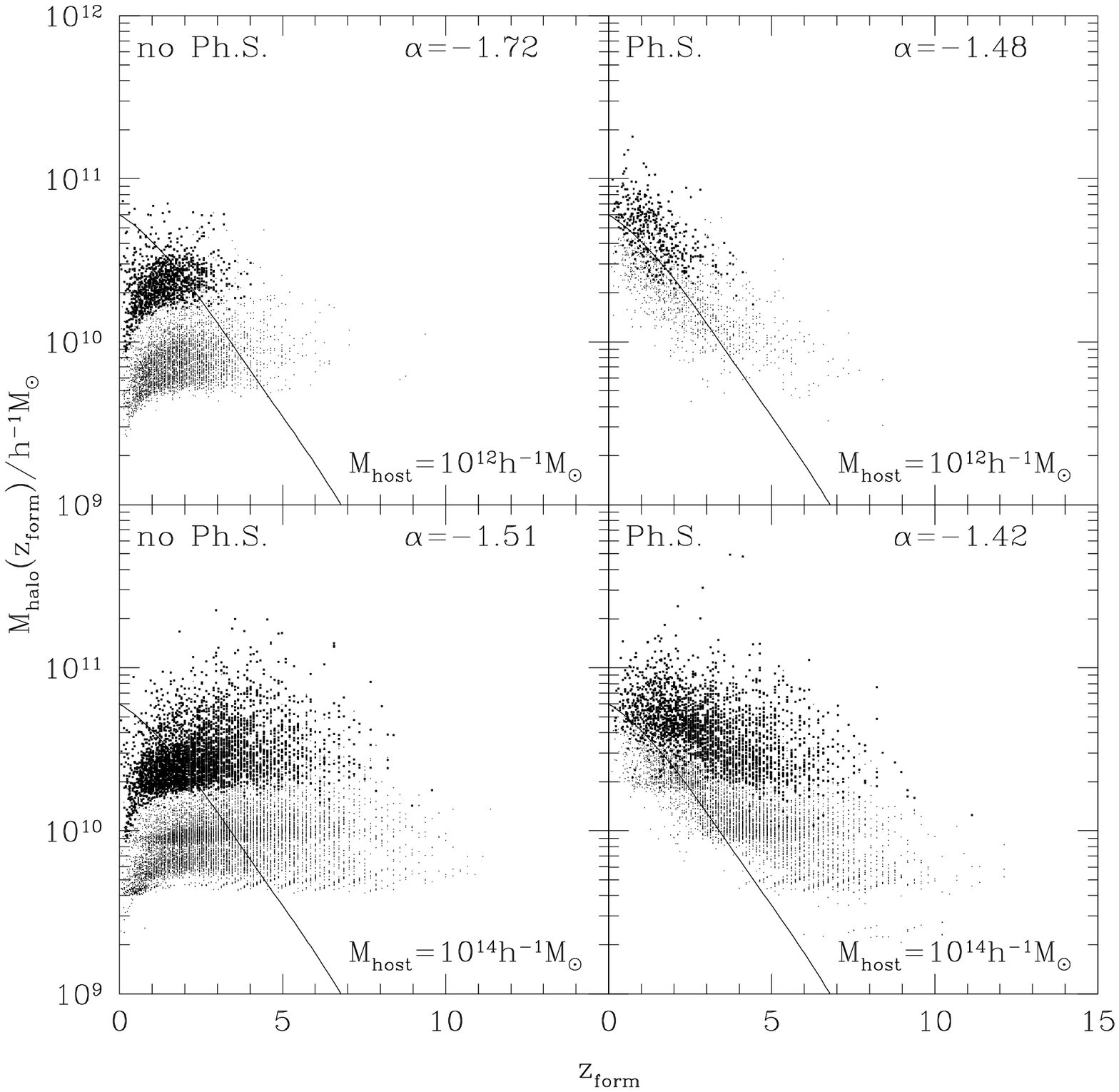,width=165mm}
\caption{Positions of galaxies in the halo-mass/formation-redshift
plane. Upper and lower panels show results for galaxies which reside
in $10^{12}h^{-1}M_\odot$ and $10^{14}h^{-1}M_\odot$ halos at $z=0$
respectively. Left and right-hand panels show results for models
ignoring \PS\ and including \PS\ respectively. The solid line shows
the filtering mass as a function of redshift. Dots indicate galaxies
in the range $-13 < M_{\rm B}-5\log h<-12$ while small squares
indicate galaxies in the range $-15 < M_{\rm B}-5\log h<-14$. An
estimate of $\alpha$ based on these two magnitude bins is shown in
each panel.}
\label{fig:explan}
\end{figure*}

These theoretical expectations are clearly manifest in
Fig.~\ref{fig:explan}, where we plot the positions of galaxies in the
halo-mass {\em vs.} formation-redshift plane. All of the galaxies we
consider at $z=0$ are satellites in a larger host halo (of mass
$M_{\rm host}$) due to the ranges of luminosity and $M_{\rm host}$ we
have chosen to consider. Therefore, we plot the formation redshift and
mass for the halo in which the satellite formed. We adopt the
definition of formation redshift of \scite{cole2000}, namely a halo is
assumed to be newly formed if its mass exceeds twice the mass of each
of its progenitors at their own formation time, and plot the halo mass
at the formation redshift (it may have increased afterwards due to
continued accretion and merging). (The formation redshifts appear
quantized in this figure because of the finite timesteps in our
calculations which, however, are sufficiently fine so as not to affect
the results.) The halo mass of the satellite is then set equal to the
mass of the halo in which it lived at redshift $z_{\rm form}$). The
position of a halo in this plane, relative to the redshift-dependent
filtering mass (shown as a solid curve), determines the degree of \PS\
it experiences. We examine faint galaxies in two luminosity ranges,
$-13 < M_{\rm B}-5\log h<-12$ (dots) and $-15 < M_{\rm B}-5\log h<-14$
(small squares).

Consider first a model with no \PS\ (left-hand panels). In this case,
faint galaxies currently hosted in a $10^{14}h^{-1}M_\odot$ cluster
halo (lower panel) have a much more extended range of formation
redshifts than galaxies in a $10^{12}h^{-1}M_\odot$ Milky Way-type
halo (upper panel). The fainter galaxies located in lower mass halos
have a more extended range of formation redshifts than their brighter
counterparts, although this difference is rather small. The values of
$\alpha$ for the two populations, given in each panel, indicate steep
slopes in both environments. The filtering mass has no influence on
these results, of course, but is shown to indicate which of halos
might be expected to still form a galaxy when \PS\ is switched on.

In the right-hand panels we plot results for a model with
\PS. The effect of the filtering mass can now be clearly
seen: halos below the solid line form galaxies much less efficiently
than before (although there are certainly still some halos that
managed to make galaxies below this line). The values of $\alpha$
given in the panels indicate that some flattening of the luminosity
function has occurred, but this effect is small.

\begin{figure*}
\psfig{file=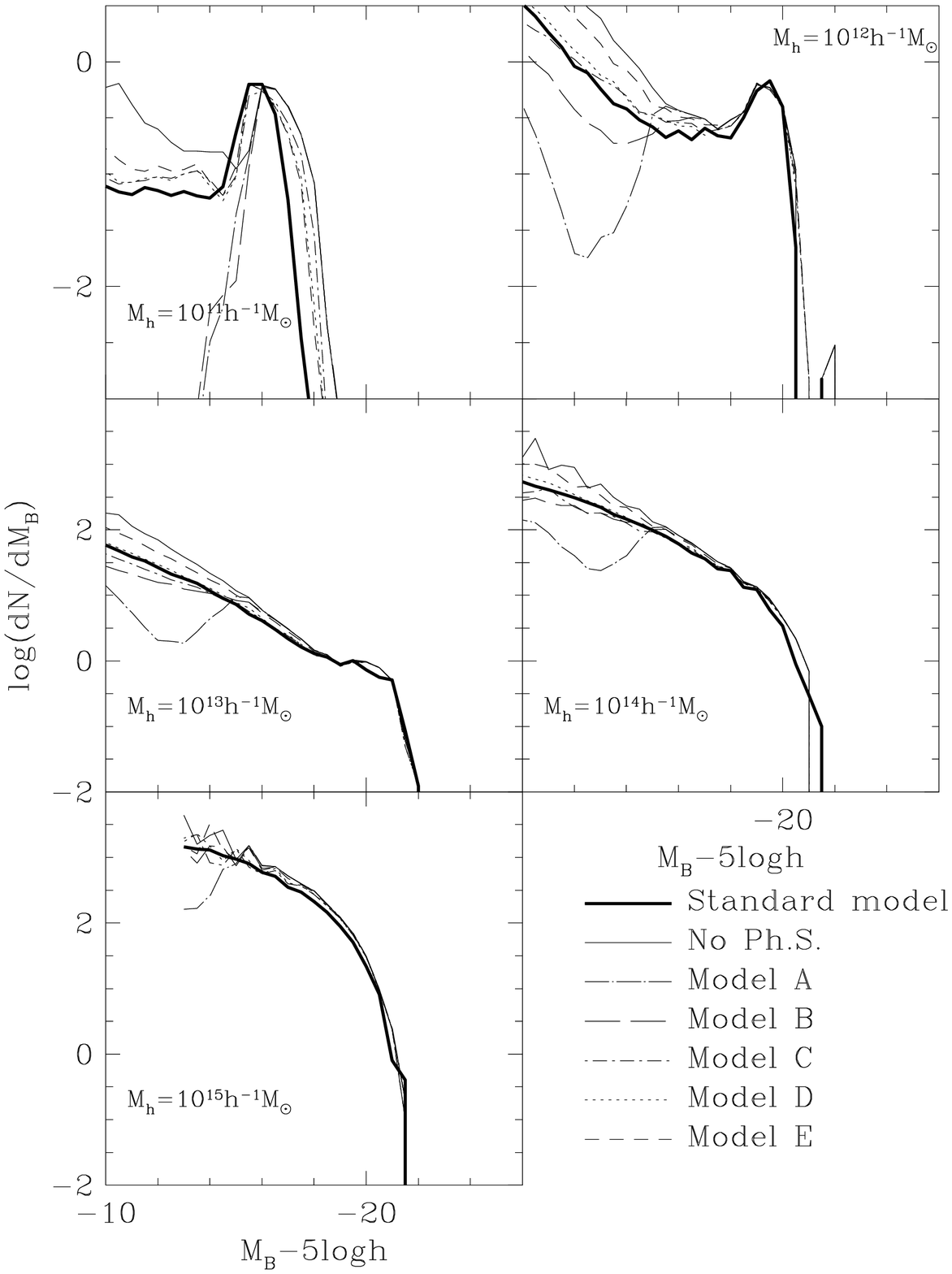,width=150mm}
\caption{B-band galaxy luminosity functions in halos of different
mass. Each panel shows the mean predicted model luminosity function in
an ensemble of dark matter halos of mass given in each panel (ranging
from halos containing small galaxies to rich clusters). All models
include the effects of tidal limitation. The heavy solid lines show
results from our complete model of \PS, while thin solid lines show a
model with no \PS. Dot-long dashed lines have a filtering mass which
jumps abruptly from zero to $3\times 10^{10}h^{-1}M_\odot$ at $z=6.5$
and a sharp suppression at that mass at lower redshifts (Model
A). Long-dashed lines have time varying filtering mass, but a sharp
suppression at that mass (Model B). The dot-short dashed line shows a
model with a fixed filtering mass but a smooth variation of
suppression with halo mass (Model C). Dotted lines are a simplified
model with time-varying filtering mass and a smooth variation of
suppression with halo mass (Model D). Short-dashed lines are the same,
but with an extra-smooth variation of suppression with halo mass
(Model E).}
\label{fig:eLFtest}
\end{figure*}

Let us now examine the importance of these various processes using
simplified models of \PS. Figure~\ref{fig:eLFtest} shows luminosity
functions in different mass halos constructed using the simplified
models described below, together with our standard model of \PS\
(heavy solid line) and a model with no \PS\ (thin solid line). 

\begin{description}
\item[{\bf Model A}] In the simplest scenario, suppose the filtering
mass jumps abruptly from zero to $M_{\rm F}$ at redshift $z_{\rm
reion}$, and that any halo of smaller mass forming at lower redshift
makes no galaxy at all, while galaxy formation in other halos proceeds
unchanged. In this case, the luminosity function should be unaffected
at bright magnitudes (since these galaxies form in halos which are
more massive than $M_{\rm F}$), should plummet sharply at a magnitude
corresponding to halo mass equal to $M_{\rm F}$, and should rise again
to fainter magnitudes (since these galaxies inhabit lower mass halos
which typically form at redshifts greater than $z_{\rm reion}$). On
the basis of eqn.~(\ref{eq:PS}), the slope of this rise should be
large for low $M_{\rm host}$ halos since these have the greatest
variation in progenitor number with mass at $z_{\rm reion}$, and
should approach some fixed value for large $M_{\rm host}$. This
behaviour is borne out when we apply this simple prescription to our
semi-analytic models, using values of $M_{\rm F}=3\times
10^{10}h^{-1}M_\odot$ (comparable to the value of $M_{\rm F}$ at the
redshifts where the majority of the faint galaxies are formed in our
standard model) and $z_{\rm reion}=6.5$ (the redshift of reionization
in our standard model; dot-long-dashed lines in
Fig.~\ref{fig:eLFtest})---suppression is greatest for small $M_{\rm
host}$, tends to steepen the faint end of the luminosity function, and
causes the most steepening for the lowest $M_{\rm host}$. This model
produces a discontinuity in the luminosity function and is a poor
approximation to the full treatment.
\item[{\bf Model B}] At the next level of sophistication, we take into
account the variation of the filtering mass with redshift as predicted
by our model, but retain the simple prescription wherein any halo
forming with mass below the filtering mass forms no galaxy at
all. Since $M_{\rm F}$ increases with time, galaxy formation in lower
mass halos is suppressed earlier, an effect which should tend to
flatten luminosity function slopes. Since halos of a given mass form
earlier in larger $M_{\rm host}$, we still expect the effects of
filtering to be weaker for clusters than for the Local Group. At a
given magnitude, this process would simply reduce the number of
galaxies. However, the detailed consequences for the faint end slope
are now harder to anticipate since they depend on just how $M_{\rm F}$
varies with redshift. Applying this prescription to our model
(long-dashed lines in Fig.~\ref{fig:eLFtest}), we still find a
discontinuity in the luminosity function (although this is much weaker
than in model (i) since $M_{\rm F}$ varies smoothly with time) which
is flattened relative to that without any \PS. Furthermore, the slopes
are significantly flatter than in case (i), indicating that the time
variation of $M_{\rm F}$ largely counteracts the steepening of the
luminosity function produced by the variation in halo formation
distributions. No environmental variation is introduced by these
processes. The resulting luminosity functions are quite similar to
those from our full \PS\ calculations, although the level of
suppression is somewhat greater. Again, the effects are largest for
the lowest $M_{\rm host}$.
\item[{\bf Model C}] Alternatively, we can consider a prescription
which has no time variation in $M_{\rm F}$, but in which, instead, the
degree of suppression varies smoothly with halo mass (as prescribed by
\pcite{gnedin00})\footnote{\protect\scite{somerville02} considered a
similar model, but kept the circular velocity corresponding to $M_{\rm
F}$ constant, resulting in a filtering mass that increased with
time. This prescription, also adopted by \protect\scite{bullock00}
provides a better match to the results of our standard model. Here,
however, we are intersted only in exploring simplified models.}. We
find that this model also produces luminosity functions similar to
those from our full \PS\ calculations (dot-short-dashed lines), and
with about the right degree of suppression for suitable choices of
$M_{\rm F}$ and $z_{\rm reion}$.
\item[{\bf Model D}] We next consider a model in which $M_{\rm F}$
varies with time and in which the degree of suppression varies
smoothly with halo mass. This then differs from our full \PS\
calculations only in neglecting suppression due to photoheating by the
ionizing background. In this model, photoionization redistributes
galaxies which previously had a given magnitude over a range of
fainter magnitudes. At a given magnitude, some galaxies are lost due
to suppression, but also new ones appear as brighter galaxies are
partially suppressed, thus becoming fainter. The net effect is
therefore difficult to judge since it depends on the time variation of
the filtering mass and on the previous shape of the luminosity
function.  We can, however, safely say that the effect is still
expected to be larger for lower $M_{\rm host}$. Applying this
prescription we reproduce the results of our full \PS\ calculation
rather well (dotted lines in Fig.~\ref{fig:eLFtest}), finding slightly
less suppression overall than in case (ii), but still with no
environmental variation.
\item[{\bf Model E}] A smoother variation of suppression with halo
mass would result in less flattening of the faint end slope. For
example, the short-dashed lines in Fig.~\ref{fig:eLFtest} show results
for a suppression of the form
\begin{equation}
M_{\rm gas} = {f_{\rm b} M_{\rm h} \over 1+M_{\rm F}/M_{\rm h}},
\end{equation}
with the same variation of filtering mass with redshift as in our
standard model.
\end{description}
In conclusion, a simple model in which galaxy formation at $z<z_{\rm
reion}$ is entirely suppressed below some fixed mass scale grossly
overestimates the effects of \PS, and produces very different
luminosity functions from those predicted by our full
calculation. Including \emph{either} a smoothly varying degree of
suppression \emph{or} a time-varying filtering mass produces better
agreement with our full calculations, indicating that these two
ingredients are of comparable importance. The degree of suppression
(as characterised by the reduction in amplitude of the LF) for a given
time dependence in $M_{\rm F}$ depends on the functional form used to
characterize the variation of suppression with halo mass. The sharper
the transition from weak to strong suppression, the greater the net
effect on the luminosity function. Smoother transitions tend to
produce steeper faint end slopes, but even the sharpest transition
possible does not produce slopes as flat as are observed and we never
find any significant environmental variation.

\end{document}